\documentstyle[11pt]{article}
\begin{document}
\begin{titlepage}
%last version
\vspace{0.2cm}

\title{ $\gamma\gamma \rightarrow t \bar{c} + c
\bar{t}$
        in a supersymmetric theory with an explicit R-parity violation
        \footnote{Supported in part by Committee of
         National Natural Science Foundation of China and Project IV.B.12 of
         scientific and technological cooperation agreement between China and
         Austria}}
\author{{ \ Yu Zeng-Hui $^{a,c}$ \ Herbert Pietschmann $^{a}$
\ Ma Wen-Gan  $^{b,c}$ \ Han Liang  $^{c}$  \ Jiang Yi  $^{c}$
 }\\\
{\small $^{a}$Institut f\" ur Theoretische Physik, Universit\" at Wien, A-1090 Vienna, Austria} \\
{\small $^{b}$CCAST (World Laboratory), P.O.Box 8730, Beijing 100080,P.R.China} \\
{\small $^{c}$Department of Modern Physics, University of Science and Technology}\\
{\small of China (USTC), Hefei, Anhui 230027, P.R.China}\\
}
\date{}
\maketitle

\vskip 12mm

\begin{center}\begin{minipage}{5in}

\begin{center} ABSTRACT\end{center}
\baselineskip 0.3in

{We studied the process $\gamma\gamma \rightarrow
t\bar{c}+c\bar{t}$ in a $R_{p}$ violating supersymmetric Model
with the effects from both B- and L-violating interactions.
The calculation shows that it is possible to detect a $R_{p}$ violating
signal at the Linear Collider. Information about the B-violating
interaction in this model
could be obtained under very clean background, if we take the present
upper bounds for the parameters
in the supersymmetric $\rlap/ R_{p}$ interactions. Even if we
can not detect a signal of $\rlap/R_{p}$ in the experiment, we
may get more stringent constraints on the heavy-flavor
$\rlap/R_{p}$ couplings.} \\

\vskip 10mm

{~~~~PACS number(s): 13.65.+i, 13.88.+e, 14.65.-q, 14.80.Dq, 14.80.Gt}
\end{minipage}
\end{center}
\end{titlepage}

\baselineskip=0.36in

\eject
\rm
\baselineskip=0.36in

\begin{flushleft} {\bf I. Introduction} \end{flushleft}
\par
  The minimal supersymmetric model (MSSM)\cite{s1} is one of the most
interesting extensions of the Standard Model (SM) and is considered as
the most favorable model beyond SM. Thus, it is interesting to confirm
whether R-parity($R_{p}$), which is introduced to guarantee the B-
and L-conservation automatically, is conserved
in the supersymmetric extension of the SM \cite{s2}.
Because of the lack of credible
theoretical arguments and experimental tests for $R_{p}$ conservation, we
can say that the $R_{p}$ violation ($\rlap/R_{p}$)would be equally well
motivated in the supersymmetric extension of the SM \cite{s3}. 
Since in the $R_{p}$-violation models supersymmetry particles can be
singly producted and neutrinos would get masses and mixing \cite{s4},
it is a significant source of new physics. Especially after the first
signals for neutrino oscillations from atmospheric neutrinos were
observed in Super-Kamiokande \cite{s5} and an anomaly 
was detected in HERA $e^{+}p$ deep
inelastic scattering(DIS) \cite{s6}, $R_{p}$-violation can 
be a good candidate to explain those experimental results.
\par
In the last few years, many efforts were made to find $\rlap/R_{p}$
interactions in experiments. Unfortunately, up to now we have only some
upper limits on $\rlap/R_{p}$ parameters, such as B-violating
$\rlap/R_{p}$ parameters ($\lambda^{''}$) and L-violating $\rlap/R_{p}$
parameters($\lambda$ and $\lambda^{'}$)\cite{s4}\cite{s7} and results
are colletec in Ref. \cite{s8} (The parameters will be defined clearly 
in the following sector). Therefore,
trying to find the signal of $R_{p}$ violation or getting more stringent
constraints on the parameters in future experiments, is a promising
task. Possible ways to find a $R_{p}$ violation signal can be
detecting odd number supersymmetric particle interactions 
as a direct signal or testing discrepancies between
predictions of $R_{p}$-conservation models and 
$R_{p}$-violation models in the experiments as indirect
informations. 
\par
In our paper we will consider the process $e^{+}e^{-}\rightarrow
\gamma\gamma
\rightarrow t \bar{c} + c \bar{t}$ in the future Linear Collider(LC).
This rare process, which is supressed by the GIM mechanism in the Standard
Model \cite{a1}, can be a good window to open new physics. In 
Ref. \cite{s9},
it was pointed out that anomalous $t\bar{q}\gamma$ coupling admitted
by present experimental results can be much larger than the prediction
of SM. Thus, $R_{p}$-violation can be a significant source which gives
this anomalous coupling.
Although small values of $\lambda^{'}$ and $\lambda^{''}$ in $\rlap/R_{p}$
theory would suppress this process, the present upper bounds on $\rlap/R_{p}$
parameters still admit experimental observation ($\lambda^{'}$ and
$\lambda^{"}$ can be of order 1 when they involve heavy
flavors, which is reasonable with assumption of family symmetry
\cite{s10}).
So we can hope that this process allows detection of $R_{p}$ violation
within the present parameter upper limits. 
\par
With the advent of new collider techniques, we can produce
highly coherent laser beams being back-scattered with high luminosity
and efficiency at the $e^{+}e^{-}$ colliders\cite{a2}.
The $\gamma \gamma$ collisions give us a very clean environment to study
the $t\bar{c}$(or $c\bar{t}$) production. Effects of L-violating parameters
in the $e^{+}e^{-}$ collisions have been
studied\cite{s7},
but only little attention was paid to B-violating parameters\cite{s11}. The
process considered here, can give us a chance to detect $B-violating$
parameters $\lambda^{''}$ in a very $clean$ environment. We can also get
information on the parameter $\lambda^{'}$ from the process, especially
for heavy flavors, which are only weakly constrained by present data.
\par
Even without $R_{p}$ violation, there are 
flavor-changing mechanisms\cite{s12} in the MSSM,
e.g.squark mixing. Therefore, $R_{p}$ violation in $\gamma\gamma
\rightarrow t\bar{c}+c\bar{t}$ can only be established if it exceeds
the value of these other mechanisms. Fortunately, in most models
with universal SUSY breaking, those contributions are small(for details
see ref. \cite{s12}). Hence we shall assume that they are suppressed
throughout our paper.  
\par
Other possible competing mechanism, such as Two-Higgs-Doublet-Model(THDM),
was considered by Atwood et al. \cite{s13} and Y.Jiang et al. \cite{s13}.
The results showed that the cross section would be much smaller assuming
the masses of higgses to be far from the c.m. energy of colliders, so
we can distinguish them easily from $R_p$-violation interactions. 
\par
In this work we concentrate on the process $e^{+}e^{-} \rightarrow
\gamma\gamma \rightarrow t \bar{c}+c \bar{t}$ in the R-parity violating
supersymmetric theory. In section 2, we give the supersymmetric
$\rlap/R_{p}$ interactions. In section 3 we give the analytical
calculations of $\gamma\gamma \rightarrow t \bar{c}+ c \bar{t}$.
In section 4 the numerical results of the process $e^{+}e^{-}
\rightarrow \gamma\gamma \rightarrow t \bar{c}+ c \bar{t}$
are presented. The conclusion is given in section 5 and some details of
the expressions are listed in the appendix.

\begin{flushleft} {\bf II. R-parity violation($\rlap/R_{p}$) in
MSSM}\end{flushleft}
\par
All renormalizable supersymmetric $\rlap/R_{p}$ interactions can
be introduced in the superpotential\cite{s8}:
$$
\begin{array} {lll}
    W_{\rlap/R_{p}} & =\frac{1}{2}
\lambda_{[ij]k} L_{i}.L_{j}\bar{E}_{k}+\lambda^{'}_{ijk}
L_{i}.Q_{j}\bar{D_{k}}+\frac{1}{2}\lambda^{''}_{i[jk]}
\bar{U}_{i}\bar{D}_{j}\bar{D}_{k}+\epsilon _{i} L_{i} H_{u}.
\end{array}
\eqno {(2.1)}
$$
where $L_i$, $Q_i$ and $H_u$ are SU(2) doublets containing lepton, quark
and Higgs superfields respectively, $\bar{E}_j$ ($\bar{D}_j$, $\bar{U}_j$)
are the singlets of lepton (down-quark and up-quark),
and $i,j,k$ are generation indices and square brackets on them denote
antisymmetry in the bracketted indices.
\par
 We ignored the last term in Eq(2.1), which will introduce 
mixing of leptons and Higgses, since  
its effects are rather small in our process  \cite{s4}\cite{14}.
So we have 9 $\lambda$-type, 27 $\lambda^{'}$-type and 9
$\lambda^{''}$-type independent parameters left. The Lagrangian density
of $\rlap/R_{p}$ is given as follows: (the lowest order of $\lambda$)
\begin{eqnarray*}
L_{\rlap/R_{p}}&=&L_{\rlap/R_{p}}^{\lambda}+L_{\rlap/R_{p}}^{\lambda^{'}}+
L_{\rlap/R_{p}}^{\lambda^{''}}
\hskip 25mm (2.2)
\end{eqnarray*}

\begin{eqnarray*}
L_{\rlap/R_{p}}^{\lambda}&=&\lambda_{[ij]k} [\tilde{\nu}_{iL}\bar{e}_{kR}e_{jL}+
\tilde{e}_{jL}\bar{e}_{kR}\nu_{iL}+
\tilde{e}^{*}_{kR}\overline{\nu_{iL}^{c}}e_{jL}-
\tilde{\nu}_{jL}\bar{e}_{kR}e_{iL}-\tilde{e}_{iL}\bar{e}_{kR}\nu_{jL}-
\tilde{e}^{*}_{kR}\overline{\nu_{jL}^{c}}e_{iL}]+h.c.
\end{eqnarray*}

\begin{eqnarray*}
L_{\rlap/R_{p}}^{\lambda^{'}}&=& \lambda^{'}_{ijk}[ \tilde{\nu}_{iL}
\bar{d}_{kR}d_{jL}+
\tilde{d}_{jL}\bar{d}_{kR}\nu_{iL}+
\tilde{d}^{*}_{kR}\overline{\nu_{iL}^{c}}d_{jL}-
\tilde{e}_{iL}\bar{d}_{kR}u_{jL}-
\tilde{u}_{jL}\bar{d}_{kR}e_{jL}-
\tilde{d}^{*}_{kR}\overline{e_{iL}^{c}}u_{jL}]+h.c.
\end{eqnarray*}

\begin{eqnarray*}
L_{\rlap/R_{p}}^{\lambda^{"}}&=&\lambda^{"}_{i[jk]}\epsilon_{\alpha\beta\gamma}
[\tilde{u}^{*}_{iR\alpha}\bar{d}_{kR\beta}d^{c}_{jR\gamma}+
\tilde{d}^{*}_{jR\beta}\bar{u}_{iR\alpha}d^{c}_{kR\gamma}+
\tilde{d}^{*}_{kR\gamma}\bar{u}_{iR\alpha}d^{c}_{jR\beta}]+h.c.
\hskip 25mm (2.3)
\end{eqnarray*}
\par
The proton lifetime suppresses the possibility of both B-violation and
L-violation, leading to the constraints:\cite{s8}
$$
|(\lambda~or~\lambda^{'}) \lambda^{"}|<10^{-10}(\frac{\tilde{m}}{100GeV})^{2}.
\eqno {(2.4)}
$$
\par
where $\tilde{m}$ is the mass of super quark or super lepton.
Therefore, we consider the contributions from
$L_{\rlap/R_{p}}^{\lambda^{'}}$
and $L_{\rlap/R_{p}}^{\lambda^{"}}$ separately.
Although the individual
parameters $\lambda$, $\lambda^{'}$ and $\lambda^{"}$ should be
typically less than
$10^{-1}-10^{-2} (\frac{\tilde{m}}{100 GeV})^{2}$\cite{s8}, we can expect
 the parameters
involving heavy flavors to be much larger in analogy with the
Yukawa couplings in the MSSM\cite{s10}. Since the constraints
on such parameters from present experimental data are rather weak,
testing $\rlap/R_{p}$ at high energy is still very important.

\begin{flushleft} {\bf III. Calculations } \end{flushleft}
\par
In the following calculations we assume the parameters $\lambda^{'}$
and $\lambda^{''}$ to be real. One-loop corrections (as shown in Fig.1)
of $\gamma(p_3)\gamma(p_4) \rightarrow t(p1)\bar{c}(p2)$
can be split into the following components:
$$
M = \delta M_{s}+ \delta M_{v} + \delta M_{b}.
\eqno {(3.1)}
$$
where $\delta M_{s}$, $\delta M_{v}$ and $\delta M_{b}$
are the one-loop amplitudes corresponding to the self-energy, vertex,
and box correction diagrams respectively. We find that amplitudes are
proportional to the products $\lambda^{'}_{i2j}\lambda^{'}_{i3j}$
($i,j=1,2,3$)
(Fig.1.(a.1-2),Fig.1.(b.1-4) and Fig.1.(c.1-8)) and
$\lambda^{"}_{2ij}\lambda^{"}_{3ij}$($i,j=1,2,3$)
(Fig.1.(a.3),Fig.1.(b.5-6) and Fig.1.(c.9-12)); thus it is possible
to detect $\rlap/R_{p}$ signals or get much stronger constraints on those
parameters by measuring this process in future LC experiments.
\par
We define the Mandelstam variables as usual
$$
    \hat{s}  = (p_{1}+p_{2})^2=(p_{3}+p_{4})^2
\eqno {(3.2)}
$$
$$
    \hat{t}  = (p_{1}-p_{3})^2=(p_{4}-p_{2})^2
\eqno {(3.2)}
$$
$$
   \hat{u}  = (p_{1}-p_{4})^2=(p_{3}-p_{2})^2
\eqno {(3.4)}
$$
The $t\bar{c}+c\bar{t}$ productions via $\gamma \gamma$ fusion
obtains contributions only from
one-loop Feynman diagrams at the lowest order.
Since the proper
vertex counterterm should cancel with the counterterms of the external
legs diagrams in this case, we do not need to deal with the ultraviolet
divergence. Thus we simply
sum over all (unrenormalized) reducible and irreducible diagrams and
the result is finite and gauge invariant. In the Appendix we will give
the details of the amplitudes. Similarly, we can get the amplitude for
subprocess $\gamma\gamma \rightarrow c \bar{t}$. Collecting all terms in
Eq.(3.1), we obtain the total cross section for the subprocess $\gamma\gamma
\rightarrow t \bar{c}+ c \bar{t}$:
\begin{eqnarray*}
\hat{\sigma}(\hat{s}) = \frac{2N_{c}}{16 \pi \hat{s}^2 }
             \int_{\hat{t}^{-}}^{\hat{t}^{+}} d\hat{t} {\bar{\sum}_{spins}^{}}
             [|M|^{2}],
~~~~~~~~~~~~~~~~~~~~~~~~~~~(3.5)
\end{eqnarray*}
where $\hat{t}^{\pm}=\frac{1}{2}\left[ (m_t^2+m_c^2-\hat{s})\pm
\sqrt{\hat{s} ^2+m_t^4+m_c^4-2\hat{s}*m_t^2-2\hat{s}*m_c^2-2
m_t^{2}*m_c^{2}} \right]$,
colour factor $N_{c}=3$ and the bar over summation means averaging over
initial spins. In order to get the observable results in the measurements of
$t \bar{c} + \bar{t} c$ production via $\gamma \gamma$ fusion
in $e^{+}e^{-}$ collider, we need to fold the cross section of
$\gamma\gamma
\rightarrow t\bar{c}+ c\bar{t}$ with the photon luminosity,
$$
\sigma(s) = \int_{(m_t+m_c)/\sqrt{s}}^{x_{max}} dz \frac{dL_{\gamma\gamma}}{dz}
              \hat{\sigma}(\hat{s}),
\eqno {(3.6)}
$$
where $\hat{s}=z^2 s$, $\sqrt{s}$ and $\sqrt{\hat{s}}$ are the $e^{+}e^{-}$
and $\gamma\gamma$ CMS energies respectively, and $\frac{dL_{\gamma\gamma}}{dz}$
is the photon luminosity, which is defined as\cite{a2}
$$
\frac{dL_{\gamma\gamma}}{dz} = 2z \int_{z^{2}/x_{max}}^{x_{max}} \frac{dx}{x}
                                F_{\gamma /e}(x)F_{\gamma /e}(z^{2}/x).
\eqno {(3.7)}
$$
The energy spectrum of the back-scattered photon is given by \cite{a2}.
$$
F_{\gamma /e}(x) = \frac{1}{D(\xi)}[1-x+\frac{1}{1-x}-\frac{4x}{\xi (1-x)}+
                   \frac{4x^{2}}{\xi^{2} (1-x)^{2}}].
\eqno {(3.8)}
$$
taking the parameters of Ref.\cite{s15}, we have
$\xi=4.8$, $x_{max}=0.83$ and $D(\xi)=1.8$.

\begin{flushleft} {\bf IV. Numerical results} \end{flushleft}
\par
In the numerical calculations, we assume
$m_{\tilde{q}}=m_{\tilde{l}}$ and consider the effects from
$L_{\rlap/R_{p}}^{\lambda^{'}}$
and $L_{\rlap/R_{p}}^{\lambda^{"}}$ separately. It will be no loss of
generality and the results could be kept in realistic models of
supersymmetry.
\par
For the B-violating parameter
$\lambda^{"}_{2ij}\lambda^{"}_{3ij}$($i,j=1-3$),
upper bounds of $\lambda^{"}_{223}$ and $\lambda^{"}_{323}$ dominate
all others,  so we will neglect all other $\lambda^{"}$ terms. For
the L-violating
parameter $\lambda^{'}_{i2j}\lambda^{'}_{i3j}$($i,j=1-3$),
we neglect all parameters except for
$\lambda^{'}_{323}$ and
$\lambda^{'}_{333}$.
\par
In Fig.2, we show the cross section of $e^{+}e^{-}\rightarrow
\gamma\gamma \rightarrow t\bar{c}+c\bar{t}$ as function of c.m.
energy of the electron-positron system at the upper bounds of
$\lambda^{"}$,
i.e. $\lambda^{"}_{323}\lambda^{"}_{223}=0.625$, see
Ref.\cite{s4}. We take $m_{\tilde{l}}=m_{\tilde{q}}=100~GeV$ (Solid
line) and $m_{\tilde{l}}=m_{\tilde{q}}=150~GeV$ (Dashed line),
respectively.
The results show that the
cross section can be $0.64~fb$ for solid line($0.29~fb$ for dashed line)
when the c.m. energy ($\sqrt{s}$) is equal to 500 GeV. So if the
electron-positron integrated luminosity of the LC is $50 fb^{-1}$, we can get
about 32 events per year when $m_{\tilde{l}}=m_{\tilde{q}}=100~GeV$.
Therefore the $\rlap/R_{p}$ signal could be detected, if
$\lambda^{''}$ are large enough under the present allowed upper bounds.
\par
In Fig.3, we plot the cross section of $e^{+}e^{-}\rightarrow
\gamma\gamma \rightarrow t\bar{c}+c\bar{t}$ as function of c.m.
energy of the electron-positron system with the upper bounds of
$\lambda^{'}$,
i.e. $\lambda^{'}_{333}\lambda^{'}_{323}=0.096$, see Ref.\cite{s4}.
We take again $m_{\tilde{l}}=m_{\tilde{q}}=100~GeV$ for solid line and
$m_{\tilde{l}}=m_{\tilde{q}}=150~GeV$ for dashed line, respectively.
The cross section is much smaller than that of
Fig.2. That looks reasonable because the upper limits
 of $\lambda^{'}$ from present data
are much smaller than those of $\lambda^{''}$.
The cross section can be only about $0.017~fb$ when $\sqrt{s}=500
GeV$,
that means we can get only 1 event per year at the LC with
integrated luminosity $50~fb^{-1}$. Thus it will be difficult to find the
signal of $\lambda^{'}$ from the process which we discussed.
\par
In order to give more stringent constraints of $\lambda^{''}$ in
future experiments, we draw the cross section at $\sqrt{s}=500~GeV$ as
function of $\lambda^{''}_{223}\lambda^{''}_{323}$ in Fig.4(Solid line is
for $m_{\tilde{l}}=m_{\tilde{q}}=100~GeV$ and dashed-line for $m_{\tilde{l}}=
m_{\tilde{q}}=150~GeV$). When
$\lambda^{''}_{223}\lambda^{''}_{323}$ is
 about 0.1, the cross section will be about
$0.02~fb$. That corresponds to 1 event per year at the LC. So if we
can't get the signal of $\rlap/R_{p}$ from the experiments, we can set
the stronger constraint on $\lambda^{''}_{223}$ and $\lambda^{''}_{323}$,
i.e. $\lambda^{''}_{223}\lambda^{''}_{323} \le 0.1$.
\par
Similarly we draw the relation between the cross section and the parameter
product $\lambda^{'}_{323}\lambda^{'}_{333}$ with $\sqrt{s}=500~GeV$
in Fig.5, solid line is for $m_{\tilde{l}}=m_{\tilde{q}}=100~GeV$ and
dashed-line for $m_{\tilde{l}}=m_{\tilde{q}}=150~GeV$.

\begin{flushleft} {\bf IV. Conclusion} \end{flushleft}
\par
We studied both
the subprocess $\gamma\gamma \rightarrow t\bar{c}+c\bar{t}$ and process
$e^{+}e^{-} \rightarrow \gamma\gamma \rightarrow t \bar{c}+c\bar{t}$ in
one-loop order in explicit $\rlap/R_{p}$ supersymmetric model.
The calculations show that we can test $\rlap/R_{p}$ theory in the
future LC experiments, if the B-violating couplings($\lambda^{''}$-type)
are large enough within the present experimentally admitted range. That
means we can detect B-violating interactions in the lepton colliders
with cleaner background. We also consider the effect from L-violating
interactions ($\lambda^{'}$-type), and conclude that it is very small in
this
process.
\par
From our calculation, we find that the subprocess
$\gamma\gamma \rightarrow t\bar{c}+c\bar{t}$ is very helpful in getting
the
information about the B-violating couplings($\lambda^{''}$). That is because
the effect of L-violating interactions($\lambda^{'}$) is small and
can be neglected. Thus if we can observe events of
this
process in the LC, we can conclude that they are from B-violation
couplings.
 Even if we can't detect any
signal from the experiments, we could improve the present upper
bounds on $\lambda^{''}_{223}\lambda^{''}_{323}$.
\par
The authors would like to thank Prof.H.Stremnitzer for reading
the manuscript.
\newpage
\begin{center} Appendix\end{center}

\par
A. Loop integrals:
\par
We adopt the definitions of two-, three-, four-point one-loop
Passarino-Veltman integral functions of reference\cite{s19}\cite{s20}.
The integral functions are defined as follows:
\par
The two-point integrals are:
$$
\{B_0;B_{\mu};B_{\mu\nu}\}(p,m_1,m_2)=
{\frac{(2\pi\mu)^{4-n}}{i\pi^2}}\int d^n q
{\frac{\{1;q_{\mu};q_{\mu}q_{\nu}\}}{[q^2-m_1^2][(q+p)^2-m_2^2]}},
~~~~~(A.a.1)
$$
The function $B_{\mu}$ should be proportional to $p_{\mu}$:
$$
B_{\mu}(p,m_{1},m_2)=p_{\mu} B_{1}(p,m_1,m_2)
~~~~~(A.a.2)
$$
Similarly we get:
$$
B_{\mu\nu}=p_{\mu}p_{\nu} B_{21}+g_{\mu\nu} B_{22}
~~~~~(A.a.3)
$$
We denote $\bar{B}_{0}= B_{0}-\Delta$, $\bar{B}_{1}= B_{1}+\frac{1}{2}\Delta$
and $\bar{B}_{21}= B_{21}-\frac{1}{3}\Delta$. with $\Delta= \frac{2}{\epsilon}
-\gamma +\log (4\pi)$, $\epsilon=4-n$. ${\mu}$ is the scale parameter.
And the three-point and four-point integrals can be obtained similarly.
\par
The numerical calculation of the vector and tensor loop integral functions
can be traced back to the four scalar loop integrals $A_0$, $B_0$, $C_0$
and $D_0$ in Ref.\cite{s19}\cite{s20} and the references therein.
\par
B. Self-energy part of the amplitude.
\par
The amplitude of self-energy diagrams $\delta M_{s}$
(Fig.1.(a)) can be decomposed into t-channel $M_{s}^{t}$ and u-channel
terms $M_{s}^{u}$. We will just give the expressions of t-channel,
and u-channel can be obtained from t-channel, changing $t$ into $u$
and exchanging all indices and arguments of the incoming photons.
The amplitude $M_{s}^{t}$ can be expressed as:
\begin{eqnarray*}
\delta M_{s}^{t} & = & \delta M_{s}^{t(a)}+
  \delta M_{s}^{t(b)}+\delta M_{s}^{t(c)}
  ~~~~~~~~~~~~~~~~~~~~~~(A.b.1)
\end{eqnarray*}
where
\begin{eqnarray*}
\delta M_{s}^{t(a)}&=& \frac{-4\pi i \alpha Q_{c}Q_{t}}
   {(t-m_{t}^{2})(t-m_{c}^{2})}  \epsilon ^{\mu} (p_3) \epsilon ^{\nu} (p_4)
   \bar{u}(p_1)\gamma_{\mu}  \\
&& (\rlap/p_{1}-\rlap/p_{3}+m_{t}) \left[ \Sigma (p_{1}-p_{3})
   \right] (\rlap/p_{1}-\rlap/p_{3}+m_{c}) \gamma_{\nu} v(p_2),
  ~~~~~~~~~~~~~~~~~~~~~~(A.b.2)
\end{eqnarray*}
\begin{eqnarray*}
\delta M_{s}^{t(b)}&=&
  \frac{-4\pi i \alpha Q_{c}^2}{(m_{t}^2-m_{c}^2)(t-m_{c}^2)}
  \epsilon ^{\mu} (p_3) \epsilon ^{\nu} (p_4) \bar{u}(p_1)
  \Sigma (p1) (\rlap/p_{1}+m_{c}) \gamma_{\mu}   \\
&& (\rlap/p_{1}-\rlap/p_{3}+m_{c}) \gamma_{\nu} v(p_2),
  ~~~~~~~~~~~~~~~~~~~~~~(A.b.3)
\end{eqnarray*}
\begin{eqnarray*}
\delta M_{s}^{t(c)}&=&
  \frac{-4\pi i \alpha Q_{t}^2}{(t-m_{t}^2)(m_{c}^2-m_{t}^2)}
\epsilon ^{\mu} (p_3) \epsilon ^{\nu} (p_4) \bar{u} (p_1)\\
&&   \gamma _{\mu} (\rlap/p_{1}-\rlap/p_{3} +m_{t})\gamma _{\nu}
  (-\rlap/p_{2}+m_{t})\Sigma (-p2)v(p_2).
~~~~~~~~~~~~~~~~~(A.b.4)
\end{eqnarray*}
where the quark electric charge $Q_{c}=Q_{t}=2/3$, $\alpha=1/137.04$ and
$\Sigma(p)$ is defined as
$$
- i \Sigma (p) ~=~ H_{L}  \rlap/p P_{L} +
          H_{R}  \rlap/p P_{R}- H^{S}_{L} P_{L} - H^{S}_{R} P_{R} \delta_{kl},
\eqno {(A.b.5)}
$$
with
$$
H_{R} ~=~ -i \Sigma _{L},
\eqno {(A.b.6)}
$$
$$
H_{R} ~=~ -i \Sigma _{R},
\eqno {(A.b.7)}
$$
$$
H^{S}_{L} ~=~ 0,
\eqno {(A.b.8)}
$$
$$
H^{S}_{R} ~=~ 0,
\eqno {(A.b.9)}
$$
where
$$
\Sigma _{L} ~=~ -\frac{i}{16 \pi ^{2}} \lambda^{'}_{i2j}\lambda^{'}_{i3j}
(B_{1}[-p,m_{q_{j}},m_{\tilde{l}_{i}}]+B_{1}[-p,m_{l_{i}},m_{\tilde{q}_{j}}]),
\eqno {(A.b.10)}
$$
$$
\Sigma_{R} ~=~ -\frac{i C_{R}}{16 \pi ^{2}}
\lambda^{''}_{2jk}\lambda^{''}_{3jk}
(B_{1}[-p,m_{q_{j}},m_{\tilde{q}_{k}}]+B_{1}[-p,m_{q_{k}},m_{\tilde{q}_{j}}]),
\eqno {(A.b.11)}
$$
where $i$ and $j,k$ are generations of leptons and quarks respectively,
$C_{R}=2$.
\par
 The amplitude from vertex diagrams and
box terms can be obtained in a similar way from Fig.1.(b,c);
however, it is very complex, so we do not express them here.
For a hint of its structure, compare with Ref.\cite{s21}

\par
%\newpage

\newpage

\begin{center}
{\large \bf Figure Captions}
\end{center}

\parindent=0pt

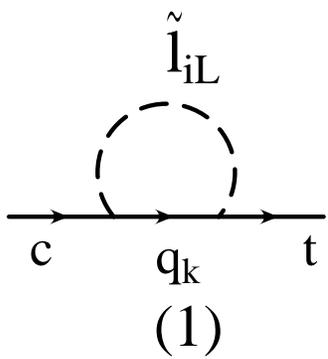
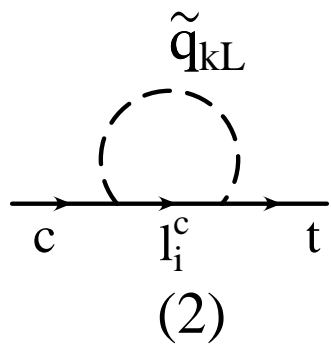
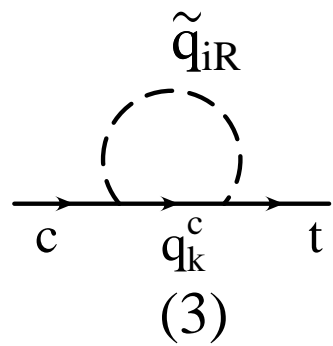
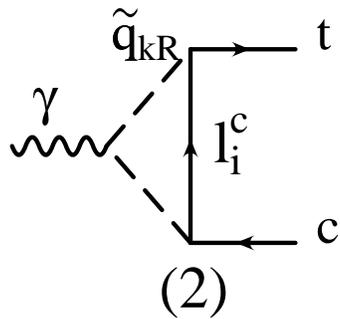
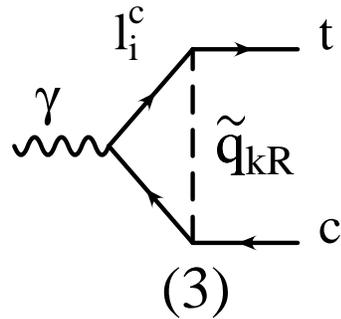
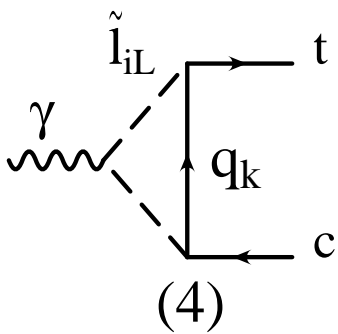
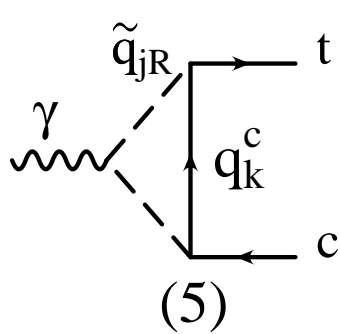
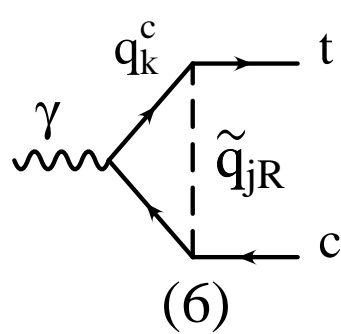
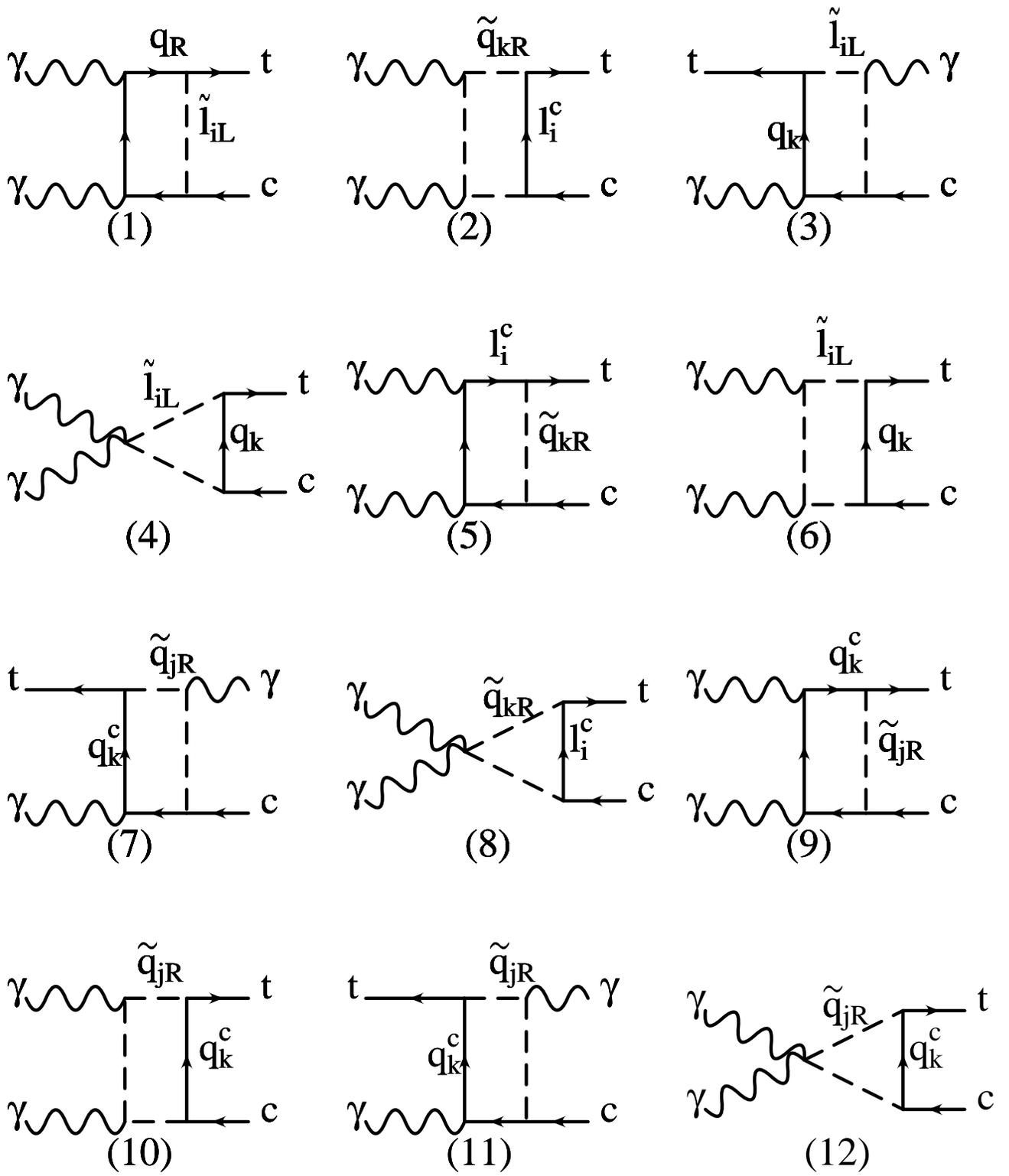
{\bf Fig.1} Feynman diagrams of $\gamma\gamma\rightarrow t \bar{c}$ subprocess.
        Fig.1 (a): Self-energy diagrams. Fig.1 (b): Vertex diagrams
        Fig.1 (c): Box diagrams (only t-channel). Dashed lines represent
        sleptons and squarks.
\par
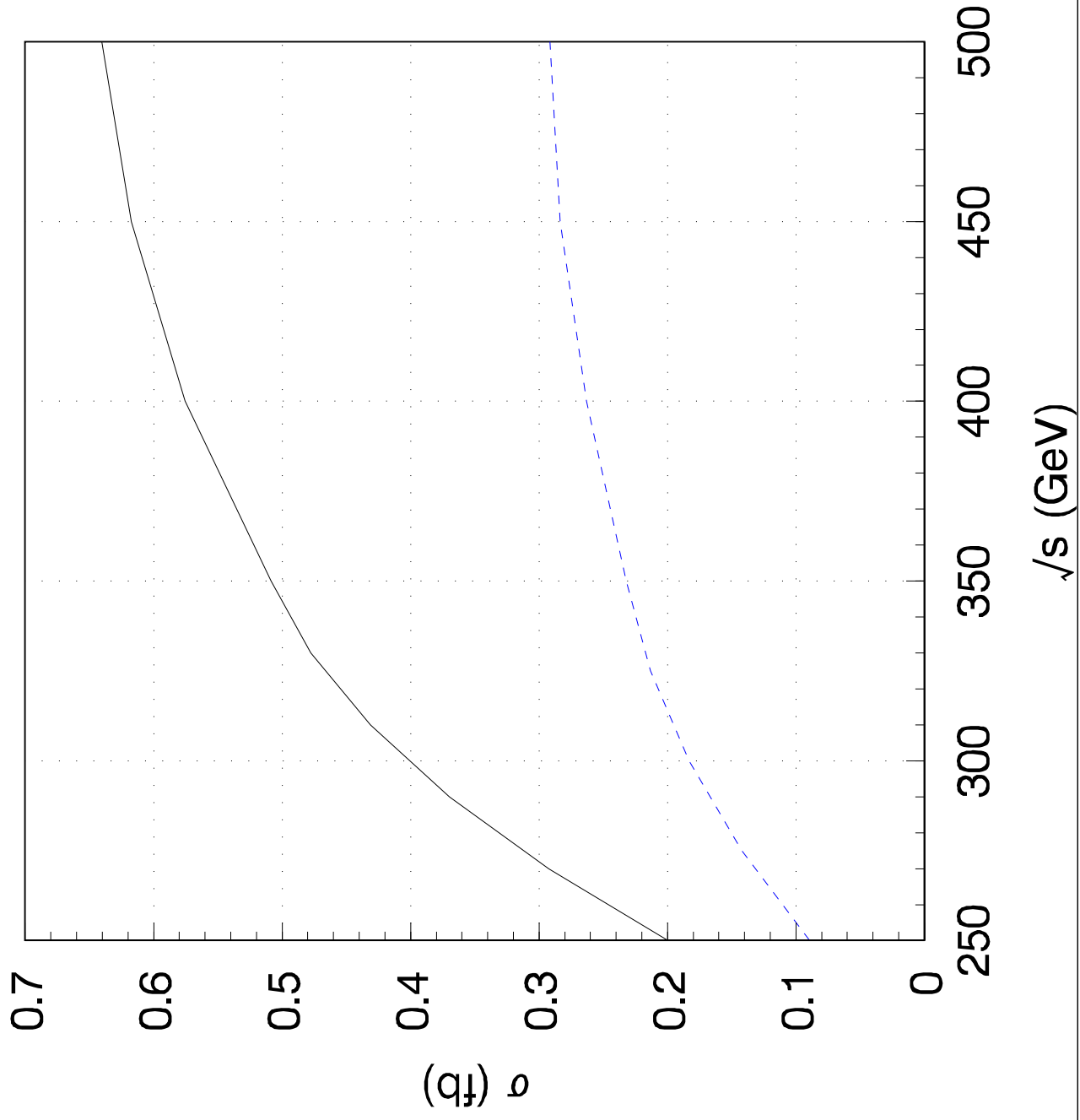
{\bf Fig.2} Cross section of $e^{+}e^{-}\rightarrow \gamma\gamma
        \rightarrow t\bar{c}+c\bar{t}$ as function of c.m.energy $\sqrt{s}$
        with $\lambda^{''}_{323}\lambda^{''}_{223}=0.625$
        solid line for $m_{\tilde{l}}=m_{\tilde{q}}=100~GeV$,
        and dashed line for $m_{\tilde{l}}=m_{\tilde{q}}=150~GeV$.
\par
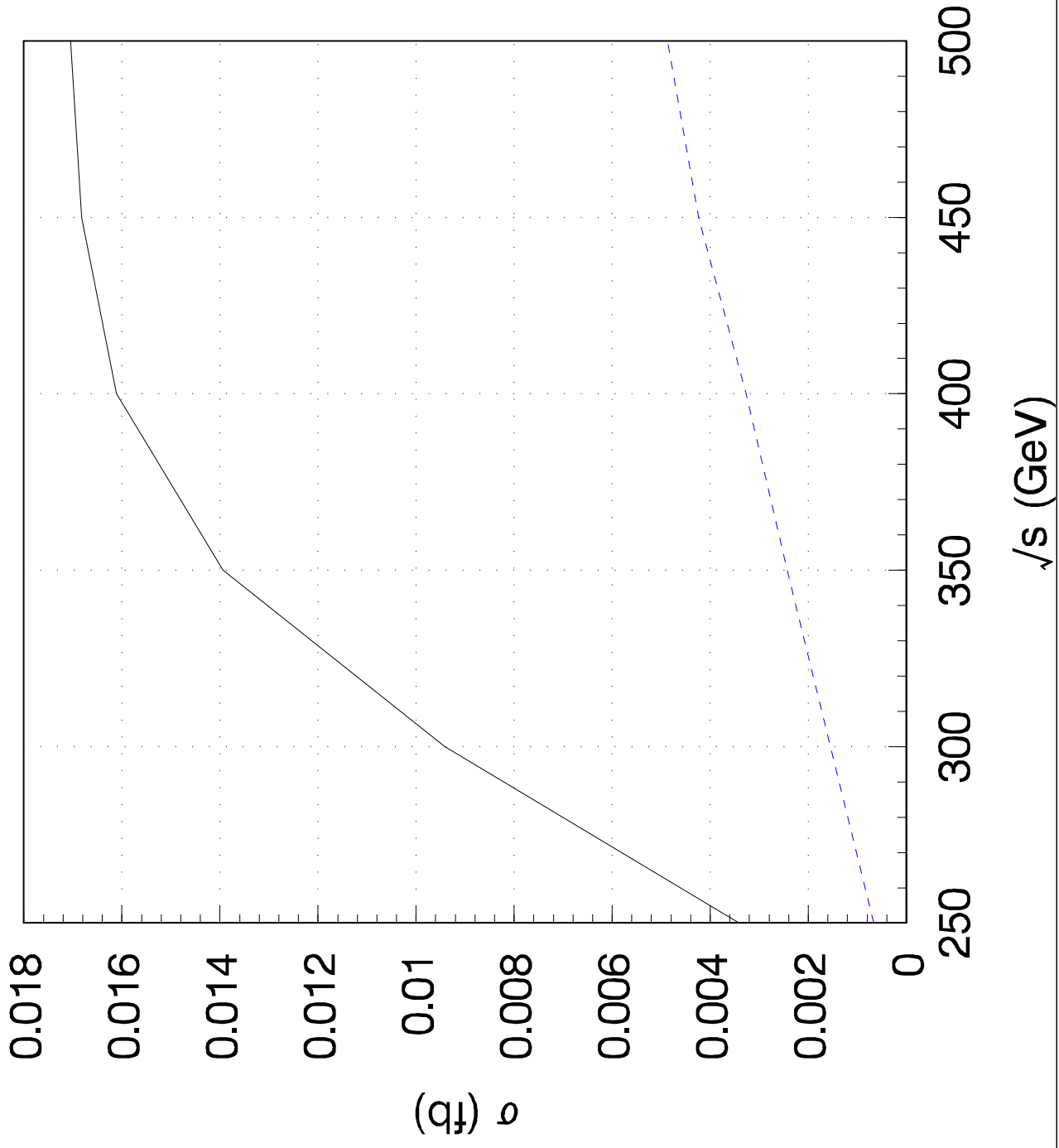
{\bf Fig.3} Cross section of $e^{+}e^{-}\rightarrow \gamma\gamma
        \rightarrow t\bar{c}+c\bar{t}$ as function of c.m.energy $\sqrt{s}$
        with $\lambda^{'}_{333}\lambda^{'}_{323}=0.096$
        solid line for $m_{\tilde{l}}=m_{\tilde{q}}=100~GeV$,
        and dashed line for $m_{\tilde{l}}=m_{\tilde{q}}=150~GeV$.
\par
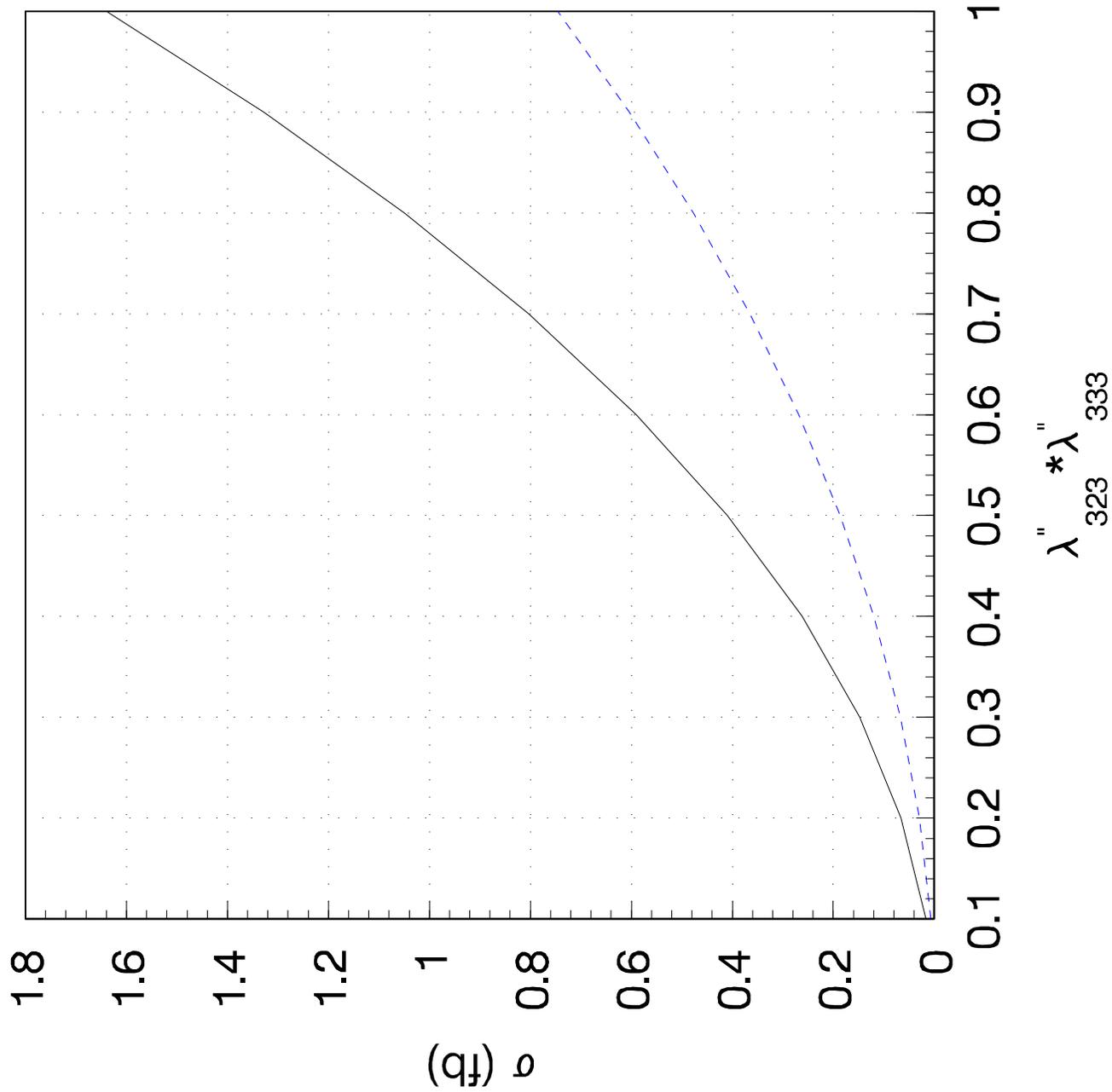
{\bf Fig.4} Cross section of $e^{+}e^{-}\rightarrow \gamma\gamma
        \rightarrow t\bar{c}+c\bar{t}$ at c.m.energy $\sqrt{s}=500~GeV$ as
        function of $\lambda^{''}_{323}\lambda^{''}_{223}$.
        solid line for $m_{\tilde{l}}=m_{\tilde{q}}=100~GeV$,
        and dashed line for $m_{\tilde{l}}=m_{\tilde{q}}=150~GeV$.
\par
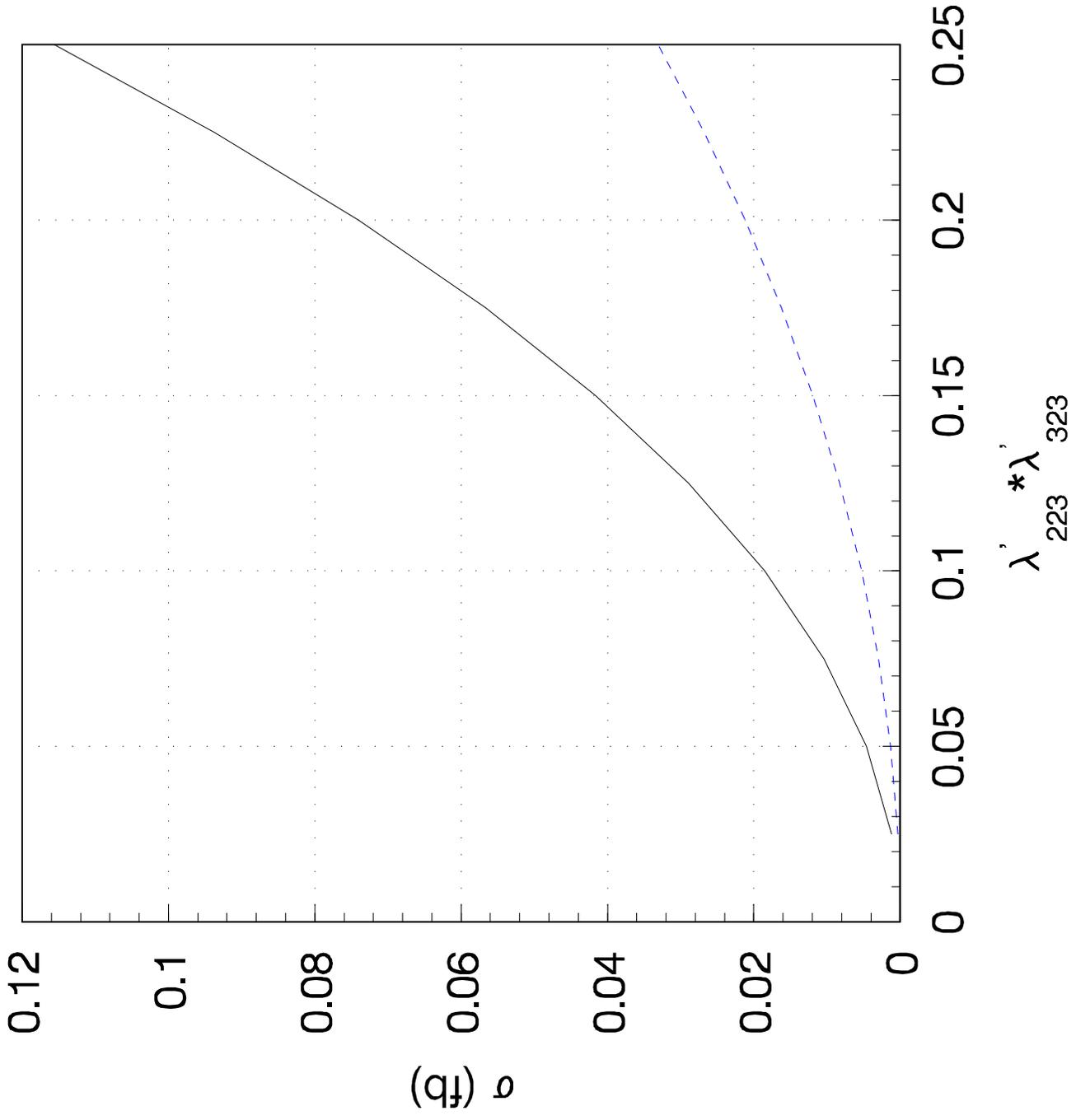
{\bf Fig.5} Cross section of $e^{+}e^{-}\rightarrow \gamma\gamma
        \rightarrow t\bar{c}+c\bar{t}$ at c.m.energy $\sqrt{s}=500~GeV$ as
        function of $\lambda^{'}_{333}\lambda^{'}_{323}$.
        solid line for $m_{\tilde{l}}=m_{\tilde{q}}=100~GeV$,
        and dashed line for $m_{\tilde{l}}=m_{\tilde{q}}=150~GeV$.
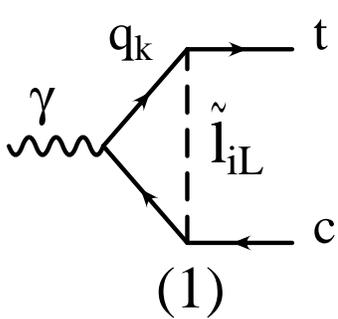
\end{document}